\documentclass[twocolumn]{aastex631}

\usepackage{graphicx}
\usepackage{xcolor}
\usepackage{MnSymbol}
\usepackage{multirow}

\begin{document}

\title{Single-epoch and Differential Astrometric Microlensing of Quasars}

\author[0000-0002-6482-2180]{R. For\'es-Toribio}
\affiliation{Departamento de Astronom\'{\i}a y Astrof\'{\i}sica, Universidad de Valencia, E-46100 Burjassot, Valencia, Spain}
\affiliation{Observatorio Astron\'omico, Universidad de Valencia, E-46980 Paterna, Valencia, Spain}

\author[0000-0003-1989-6292]{E. Mediavilla}
\affiliation{Instituto de Astrof\'{\i}sica de Canarias, V\'{\i}a L\'actea S/N, La Laguna, E-38200, Tenerife, Spain}
\affiliation{Departamento de Astrof\'{\i}sica, Universidad de la Laguna, La Laguna, E-38200, Tenerife, Spain}

\author[0000-0001-9833-2959]{J. A. Mu\~noz}
\affiliation{Departamento de Astronom\'{\i}a y Astrof\'{\i}sica, Universidad de Valencia, E-46100 Burjassot, Valencia, Spain}
\affiliation{Observatorio Astron\'omico, Universidad de Valencia, E-46980 Paterna, Valencia, Spain}

\author[0000-0001-7798-3453]{J. Jim\'enez-Vicente}
\affiliation{Departamento de F\'{\i}sica Te\'orica y del Cosmos, Universidad de Granada, Campus de Fuentenueva, 18071 Granada, Spain}
\affiliation{Instituto Carlos I de F\'{\i}sica Te\'orica y Computacional, Universidad de Granada, 18071 Granada, Spain}

\author[0000-0002-2306-9372]{C. Fian}
\affiliation{Departamento de Astronom\'{\i}a y Astrof\'{\i}sica, Universidad de Valencia, E-46100 Burjassot, Valencia, Spain}

\author[0000-0002-8949-5200]{C. del Burgo}
\affiliation{Instituto de Astrof\'{\i}sica de Canarias, V\'{\i}a L\'actea S/N, La Laguna, E-38200, Tenerife, Spain}
\affiliation{Departamento de Astrof\'{\i}sica, Universidad de la Laguna, La Laguna, E-38200, Tenerife, Spain}

\correspondingauthor{Raquel For\'es-Toribio} \email{raquel.fores@uv.es}

\begin{abstract}
We propose and discuss a new experimental approach to measure the centroid shift induced by gravitational microlensing in the images of lensed quasars (astrometric microlensing). Our strategy is based on taking the photocenter of a region in the quasar large enough as to be insensitive to microlensing as reference to measure the centroid displacement of the continuum. In this way, single-epoch measurements of astrometric microlensing can be performed. Using numerical simulations, we show that, indeed, the centroid shift monotonically decreases as the size of the emitting region increases, and only for relatively large regions, like the broad line region (BLR), does the centroid shift become negligible. This opens interesting possibilities to study the stratification of the different emitters in the accretion disk and the BLR. We estimate the amplitude of the centroid shifts for 79 gravitationally lensed images and study more thoroughly the special cases Q2237+030 A, RXJ1131$-$1231 A, PG1115+080 A2 and SDSS~J1004+4112 A. We propose to use spectro-astrometry to simultaneously obtain the photocenters of the continuum and of different emission line regions since, with the precision of forthcoming instruments, astrometric microlensing by $\sim 1 M_\odot$ mass microlenses may be detected in many quasar lensed images. When we consider more massive micro/millilenses, $M\gtrsim 10 M_\odot$, often proposed as the constituents of dark matter, the BLR becomes sensitive to microlensing and can no longer be used as a positional reference to measure centroid shifts. Differential microlensing between the images of a lensed quasar along several epochs should be used instead.
\end{abstract}

\section{Introduction \label{intro}}

By studying the effect of gravitational microlensing on lensed quasars \citep{1979Natur.282..561C,2006glsw.conf..453W}, it is possible to infer properties concerning both the microlenses and the quasars: whether there are compact objects within the lens and, if so, what is their mass distribution \citep[see, e.g.,][]{2004ApJ...605...58K,2009ApJ...706.1451M,2019ApJ...885...75J,2023A&A...673A..88A}, and the size and brightness profile of the quasars \citep[see, e.g.,][]{2008A&A...490..933E,2012ApJ...755...82M,2012ApJ...756...52M,2014ApJ...783...47J,2016ApJ...817..155M,2023A&A...678A.108F}. In order to draw conclusions about these properties, the classical approach is to measure temporal changes in the brightness of the lensed images. 

Microlensing-induced changes in brightness are caused by the splitting of a macroimage in several deformed and irregularly disposed microimages that cannot be resolved with present-day telescopes, but affects the centroid of the observed macroimages. Thus, present and future improvements in astrometric instrumentation may allow to study the astrometric microlensing effect. The main idea is to measure the photocenter of the images and study their shifts, instead of their temporal variability in brightness. \citet{1998ApJ...501..478L} first proposed this type of study in the context of galaxy-quasar microlensing. \citet{2002ApJ...568..717H,2003AJ....125.1033S,2004A&A...416...19T,2007ApJ...659...52C,2013MNRAS.432..848P} continued to explore the astrometric microlensing and millilensing of quasars in different scenarios.

A possible baseline for the displacements can be a theoretical reference obtained from a macromodel of the lens system. However, the models can have systematic positional errors of around several milliarcseconds (mas), see e.g., \citet{2007ApJ...659...52C,2012A&A...538A..99S}. Another procedure proposed by \citet{2004A&A...416...19T} consists of measuring the relative shift between multiple images of a lensed quasar. The main drawback of this approach is that it relies on measurements over months or years as the classical method to study microlensing based on changes in magnitude.

Here, we are going to propose and study a third approach. It consists of choosing a larger component of the quasar as the baseline. The principal hypothesis is that the larger component is less affected by the microlensing effect since the distortions in the final image produced by the many stars back-mapped to the region covered by the source at the source plane are compensated and diluted. Thus, the first step to apply the single-epoch method is to show that the displacement produced by the stars decreases with increasing size. The most important advance of this method is that it can be applied to single-epoch measurements since their reference does not depend on previous measurements. 

Another basic information needed to asses the feasibility of astrometric microlensing measurements in different scenarios is the amplitude of the centroid shifts. To obtain an order of magnitude estimate we can consider that for typical fractions of mass in stars ($\alpha \sim 10-20$ \%) and sizes of the quasar source (from $1$ to $5$ light-days), the rays coming from the source will be mainly affected by a single or a few stars (low optical depth approximation) and, consequently, we can expect that the images of the source are going to be asymmetrically distributed with a separation of the order of an Einstein radius. This separation is going to be enlarged by the macromagnification induced by the total distribution of matter (microlenses and smooth component) and the external shear. Thus, the centroid shift along each principal axis would be of the order of an Einstein radius multiplied by some linear magnification (to be determined with simulations). Then, if the linear magnification is $\mu_\pm$ along each principal axis (as defined in \S\ref{sec:method}), respectively, we can expect a centroid shift of around $\sqrt{\mu_+^2+\mu_-^2}$ Einstein radii. For a $0.3M_\odot$ star the Einstein ring of a typical lens (source at redshift 2 and lens at redshift 0.5), is $10$ light-days at the source plane, which corresponds to 1 microarcsecond, $\mu$as. Taking into account that the Einstein radius is proportional to $\sqrt{M}$, we can expect centroid shifts of, roughly, 
\begin{equation}
\label{estima}
\Delta C \sim \sqrt{\mu_+^2+\mu_-^2}\sqrt{M/0.3M_\odot}\, \rm \mu as. 
\end{equation}
Hence, centroid shifts of a few tens of $\mu$as, measurable with existent and planned instrumentation, are expected. If we consider microlenses more massive than the stars, the centroid shifts can be as large as hundreds of $\mu$as (for LIGO-Virgo-KAGRA detected black holes) or several mas (for primordial massive black holes, PMBH, or primordial black hole clusters with masses larger than $10^3M_\odot$). However, for lenses masses above the stellar mass range, the original reference can also experiment centroid shifts and we should either look for a larger region unaffected for these mass ranges (like the narrow line region, NLR) or, alternatively, measure the differential centroid shifts between lensed quasar images in several epochs.

The article is organized as follows. In \S\ref{sec:method} we describe the methodology of the numerical simulations. The dependence of the centroid shift with several variables for a generic lens system is studied in \S\ref{sec:generic}, and \S\ref{sec:specific} is devoted to a quantitative study of the effect for specific lens systems. Possible experimental measurements in different scenarios are discussed in \S\ref{sec:disc}, and the main conclusions are summarized in \S\ref{sec:concl}.

\section{Numerical methods and simulations}\label{sec:method}

We perform all simulations with an adaptation of the Fast Multipole Method – Inverse Polygon Mapping\footnote{https://gloton.ugr.es/microlensing/} (FMM-IPM) developed by \citet{2022ApJ...941...80J}. We calculate the image of a source by inverse ray shooting. As there are many lenses and rays involved, we use the FMM to effectively calculate the ray's deflections.

We model the sources as uniform circles sampled with at least $10^5$ pixels regardless of the source size. Similarly, the resolution in the image plane is set so that all images are sampled with at least N=$10^5$ pixels in absence of microlensing. The area of the imaged source with radius $r_s$ in units of the Einstein radius, $R_E$, in an environment of macromagnification $\mu$ is $A=\mu\pi r_s^2$. The resolution (res=$R_E/$pixel) that makes this area sampled by at least N pixels is $A/\text{res}^2\geq N$ or equivalently $\text{res}\leq\sqrt{\mu\pi r_s^2/N}$. In this way, all images are, on average, sampled with the same number of pixels. Given that the stochastic deflection of the stars is $\sqrt{\kappa_\ast}R_E$ \citep{1986ApJ...306....2K,2021arXiv210412009D}, the image plane size needs to be enlarged beyond the magnification in each principal axis, i.e., $\mu_-=1/(1-\kappa-\gamma)$ and $\mu_+=1/(1-\kappa+\gamma)$, to ensure that the majority of the microimages are recovered. \citet{2022ApJ...931..114Z} defined a ``protective border" of $10\sqrt{\kappa_\ast}R_E$ per side, which ensures that at least 98\% of the light rays fall into the studied region of the source plane. In our case, the size of the image plane is set to the magnification in each principal axis multiplied by an ``allowance" factor. This factor is determined after a preliminary study for each individual case for which the image centroid is stabilized and always enlarges the image plane area more than $10\sqrt{\kappa_\ast}R_E$ per side. Generally, we need to extend the image plane beyond that border limit because accurate centroid determinations require the retrieval of microimages located far from the image plane center.

The stars have the same mass and are randomly distributed in a circle on the image plane. The number of stars is fixed in the simulations for each study case (i.e., Sections \ref{sec:gen_mag}, \ref{sec:gen_size-alpha}, \ref{sec:sp_2237}, \ref{sec:sp_1131}, \ref{sec:sp_1115}, and \ref{sec:sp_1004}) to keep the same Poissonian uncertainty. The number of stars needed to stabilize the centroid shift in the worst-case scenario of each section is the one adopted for the rest of the simulations. To achieve sufficient statistical significance, all results are reported after simulating 1000 realizations for each set of parameters.

We use as an input the convergence $\kappa$, shear $\gamma$, fraction in stars $\alpha=\kappa_{\ast}/\kappa$, as well as the stellar distribution and the source radius 
$r_s$. We calculate the image of our circular uniform source and save the pixels (j1,j2) in the image plane where there is an image (i.e. the inverse ray hits the source). For the rest of the image plane we have I(j1,j2)=0. From this image matrix $I$, we can compute the image centroid in image plane pixel coordinates as:
\begin{equation}
j_{\text{cent},k} = \frac{\sum\limits_{j_1=1}^{n_{1}} \sum\limits_{j_2=1}^{n_{2}} I(j_1,j_2)j_k}{\sum\limits_{j_1=1}^{n_{1}} \sum\limits_{j_2=1}^{n_{2}} I(j_1,j_2)} ,
\end{equation}
with $k$=1, 2 for the principal axes and $n_1$, $n_2$ are the number of pixels in each image plane axis. The centroid in pixel coordinates $(j_{\text{cent},1},j_{\text{cent},2})$ is transformed into physical coordinates $(\Delta C_1,\Delta C_2)$, whose origin is at the center of the source.
 
\section{Results for a generic case}\label{sec:generic}

To better grasp the qualitative phenomenology of astrometric microlensing we start studying a generic case fixing as few parameters of the lens system as possible. The idea is to express all the computations in terms of the Einstein radius, which encompasses the geometry of the system (i.e. the redshifts of the source and the lens, $z_s$ and $z_l$, respectively) and the mass of the microlenses, $M$.

\begin{figure}
\includegraphics[width=\linewidth]{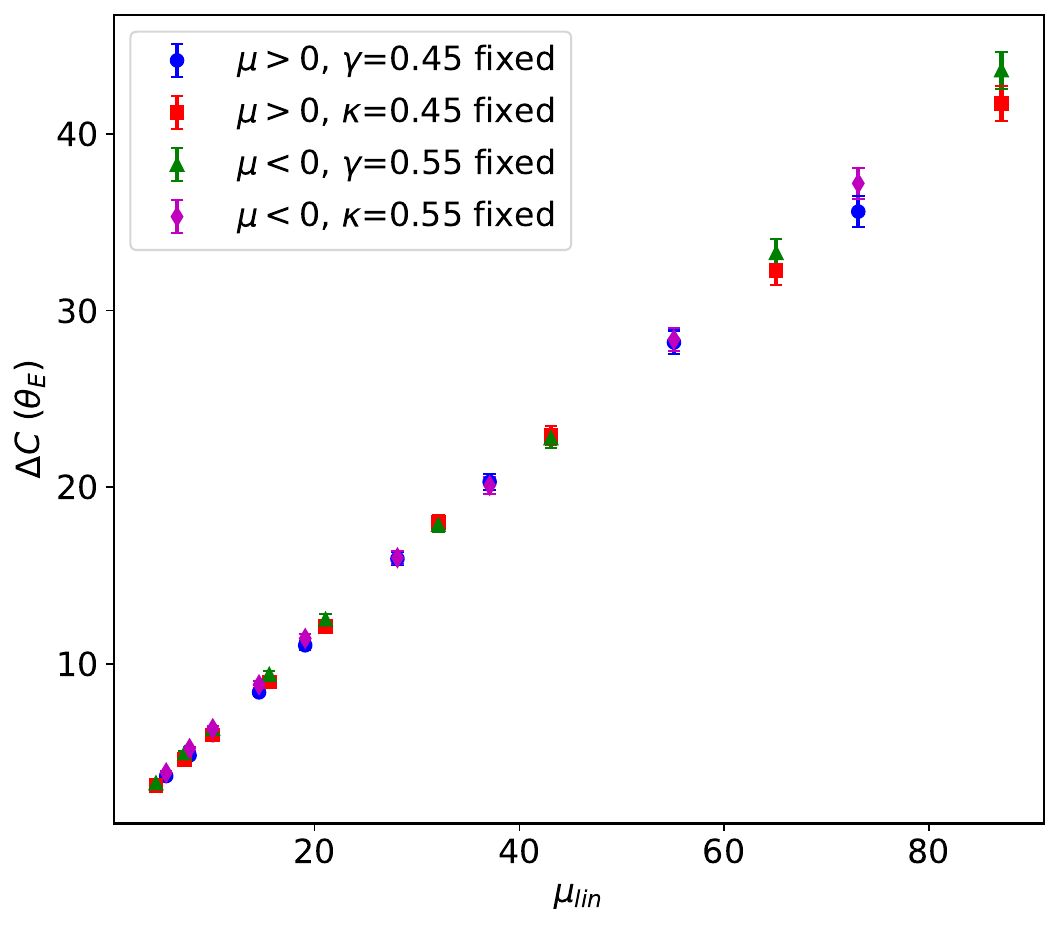}
\caption{Centroid shift dependence of a 1$R_E$ source for different values of convergence and shear while keeping the convergence in microlenses fixed to $\kappa_{\ast}$=0.09 with respect to the linear macromagnification, $\mu_{\text{lin}}$. \label{fig:shift_mag}}
\end{figure}

\begin{figure*}
\includegraphics[width=0.5\linewidth]{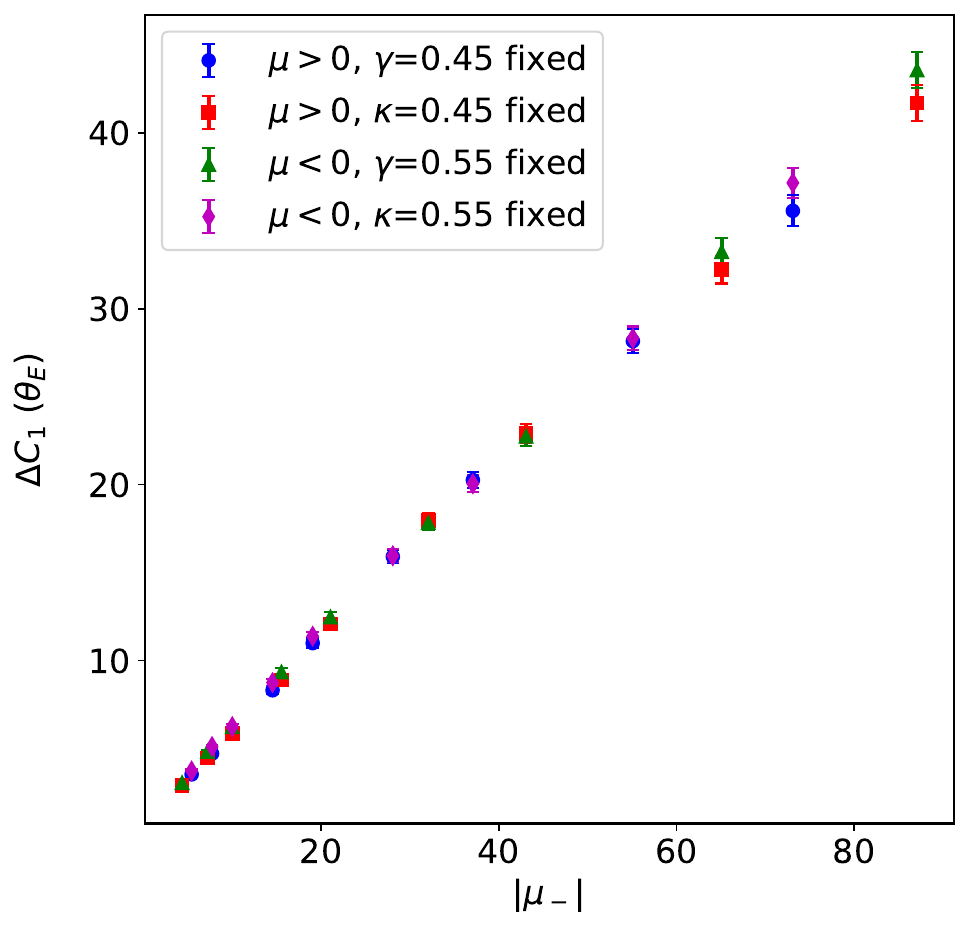}
\includegraphics[width=0.5\linewidth]{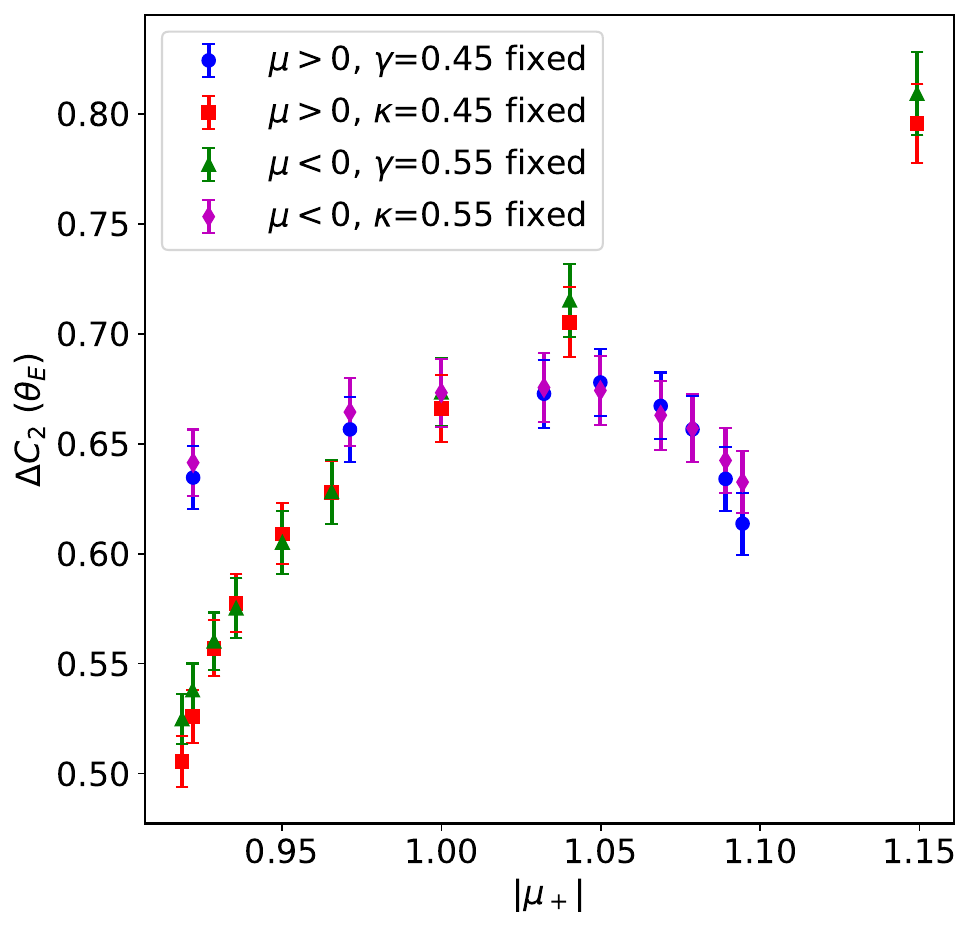}
\caption{Centroid shift dependence of a 1$R_E$ source in each principal axis ($\Delta C_1$ and $\Delta C_2$) for different values of convergence and shear while keeping the convergence in microlenses fixed to $\kappa_{\ast}$=0.09. Each centroid shift is represented against the magnification in the corresponding axis, $|\mu_-|$ and $|\mu_+|$ respectively. \label{fig:shift_axes}}
\end{figure*}

\subsection{Dependence of the centroid shift with the linear macromagnification}\label{sec:gen_mag}

We study the centroid displacement of a source of $r_s=1R_E$ caused by a population of stars with convergence $\kappa_{\ast}=0.09$ in different macromagnifications environments. This convergence in microlenses corresponds to $\alpha=0.2$ for a typical convergence value of $\kappa=0.45$, which is the average stellar fraction in lens galaxies \citep{2015ApJ...799..149J}. We start from the typical values $\kappa=\gamma=0.45$ ($\mu=+10$) for positive magnification and $\kappa=\gamma=0.55$ ($\mu=-10$) for negative magnification. Then, in each case we vary either $\kappa$ (without changing $\kappa_{\ast}$) or $\gamma$ to explore the magnification interval from $\pm5$ to $\pm80$. 

In Figure \ref{fig:shift_mag}, we represent the dependence of the centroid shifts, $\Delta C=\sqrt{\Delta C_1^2+\Delta C_2^2}$, with the linear macromagnification defined as:
\begin{equation}
\mu_{\text{lin}}\coloneq\sqrt{\mu_-^2+\mu_+^2}=\sqrt{\frac{1}{(1-\kappa-\gamma)^2}+\frac{1}{(1-\kappa+\gamma)^2}}.
\end{equation}

According to this trend, the larger the linear macromagnification, the larger the centroid shifts, as expected, regardless of whether $\kappa$ or $\gamma$ is varied or whether positive or negative magnification is considered. This is due to the fact that the centroid shift is the square root of the quadratic sum of the shift in each axis, which in turn can be expressed as the shift produced only by the stars, $\Delta \vec C_0$, multiplied by the magnification of each principal axis:
\begin{equation}\label{eq:Delta_vec}
\Delta \vec C =  \left(\begin{array}{cc} \frac{1}{1 -\kappa-\gamma} & 0 \\ 0 & \frac{1}{1-\kappa+\gamma} \\ \end{array}\right)\Delta \vec C_0.
\end{equation}

In absence of shear, the shifts produced by the stars in each axis are symmetric \citep{1986ApJ...306....2K} and we can express the centroid shift as the displacement in absence of magnification multiplied by the linear macromagnification:
\begin{equation}\label{eq:Delta_mu}
\Delta C=\sqrt{\frac{\Delta C_{0,1}^2}{(1-\kappa-\gamma)^2}+\frac{\Delta C_{0,2}^2}{(1-\kappa+\gamma)^2}}\simeq \frac{1}{\sqrt{2}}\, \Delta C_0 \, \mu_{\text{lin}},
\end{equation}
with $\Delta C_0=\sqrt{\Delta C_{0,1}^2+\Delta C_{0,2}^2} \simeq \sqrt{2}\, |\Delta C_{0,1}|$ as $|\Delta C_{0,1}|\simeq|\Delta C_{0,2}|$. If the displacement produced by the stars is constant, then a linear relation is established between $\Delta C$ and $\mu_{\text{lin}}$.

However, we can see a decrease in the slope of Figure \ref{fig:shift_mag} as $\mu_{\text{lin}}$ grows larger, even though the convergence in stars is kept fixed. To explore this behavior, we show in Figure \ref{fig:shift_axes} the contribution to the centroid shifts from each axis against the corresponding magnification, $\mu_-$ and $\mu_+$ in absolute value. The total centroid shift is dominated by the contribution from axis 1 which has substantially larger magnifications than axis 2. For the latter direction, the covered range of magnifications is so small that we cannot consistently recover the linear relation. The expected increasing trend for the cases ($\mu>0$, $\gamma$=0.45) and ($\mu<0$, $\kappa$=0.55) is broken for larger $|\mu_+|$ which correspond to larger $\mu_{\text{lin}}$ and the deviation from linearity for the cases ($\mu>0$, $\kappa$=0.45) and ($\mu<0$, $\gamma$=0.55) occur at smaller $|\mu_+|$ which also belong to the cases of large $\mu_{\text{lin}}$. The decrease in the slope of $\Delta C_1$ is also present at around $|\mu_-|\sim 60$. This may be related to an increase in the number of stars significantly contributing to the microlensing of the source (departure from the low optical depth approximation). The number of microlenses in an area of the image plane, $A_x$, back-mapped to an area $A_y$ in the source plane is given by
\begin{equation}
N_{\ast}=\frac{\kappa_{\ast}}{\pi}|\mu|A_y.
\end{equation}
If we consider that $A_y$ subtends the area corresponding to a circular source size, we obtain $N_{\ast}=\kappa_{\ast}|\mu|r_s^2$. For the typical case, $\kappa_*=0.09$, $\mu=10$ and $r_s=1R_E$, there are $N_{\ast}=0.9$ stars in the area of the source. According to Figure \ref{fig:shift_mag}, the departure from linearity is notorious when, on average, $\sim5$ stars in the source area ($|\mu|\sim60$) are present.

\begin{figure*}
\includegraphics[width=0.5\linewidth]{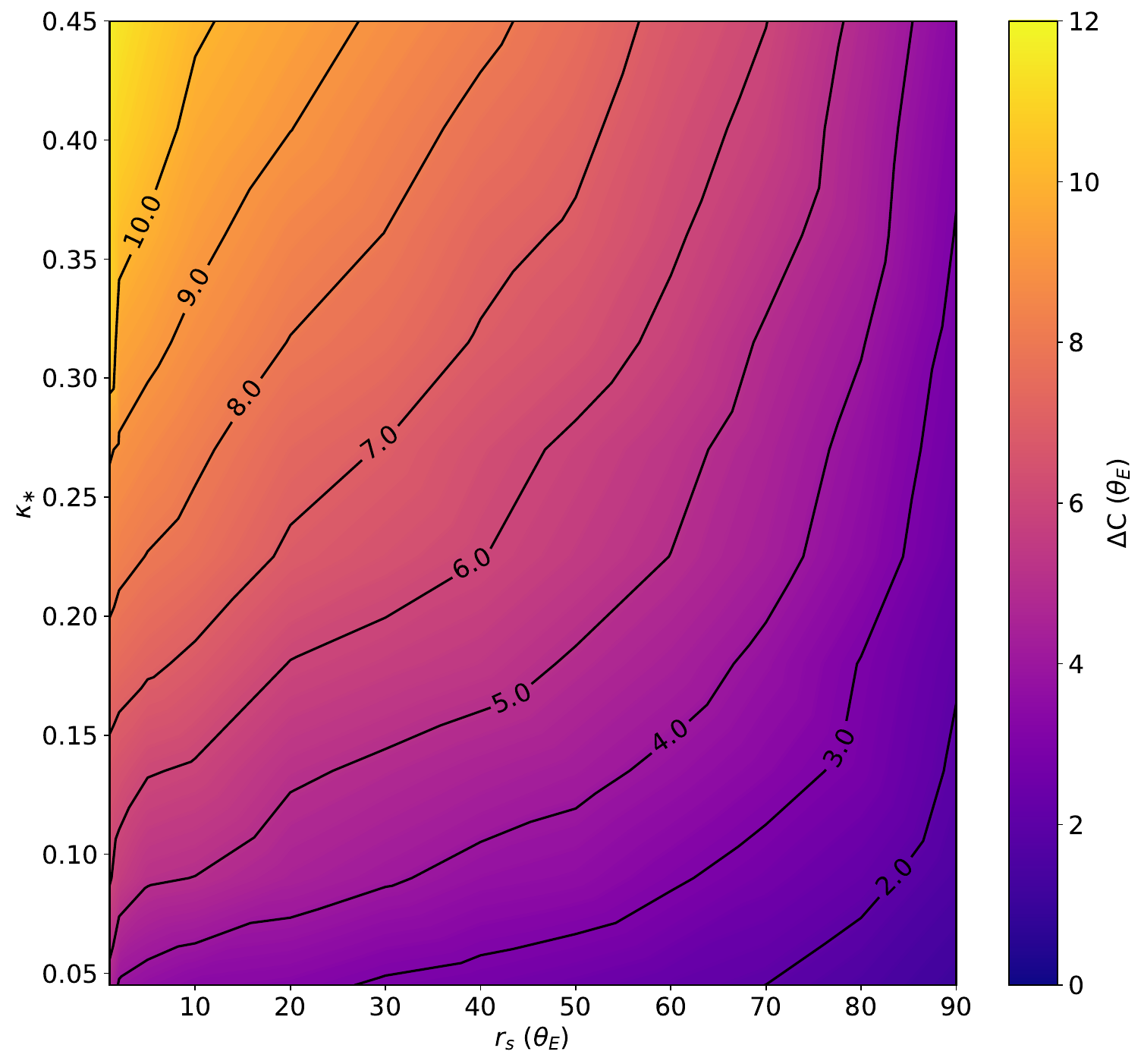}
\includegraphics[width=0.5\linewidth]{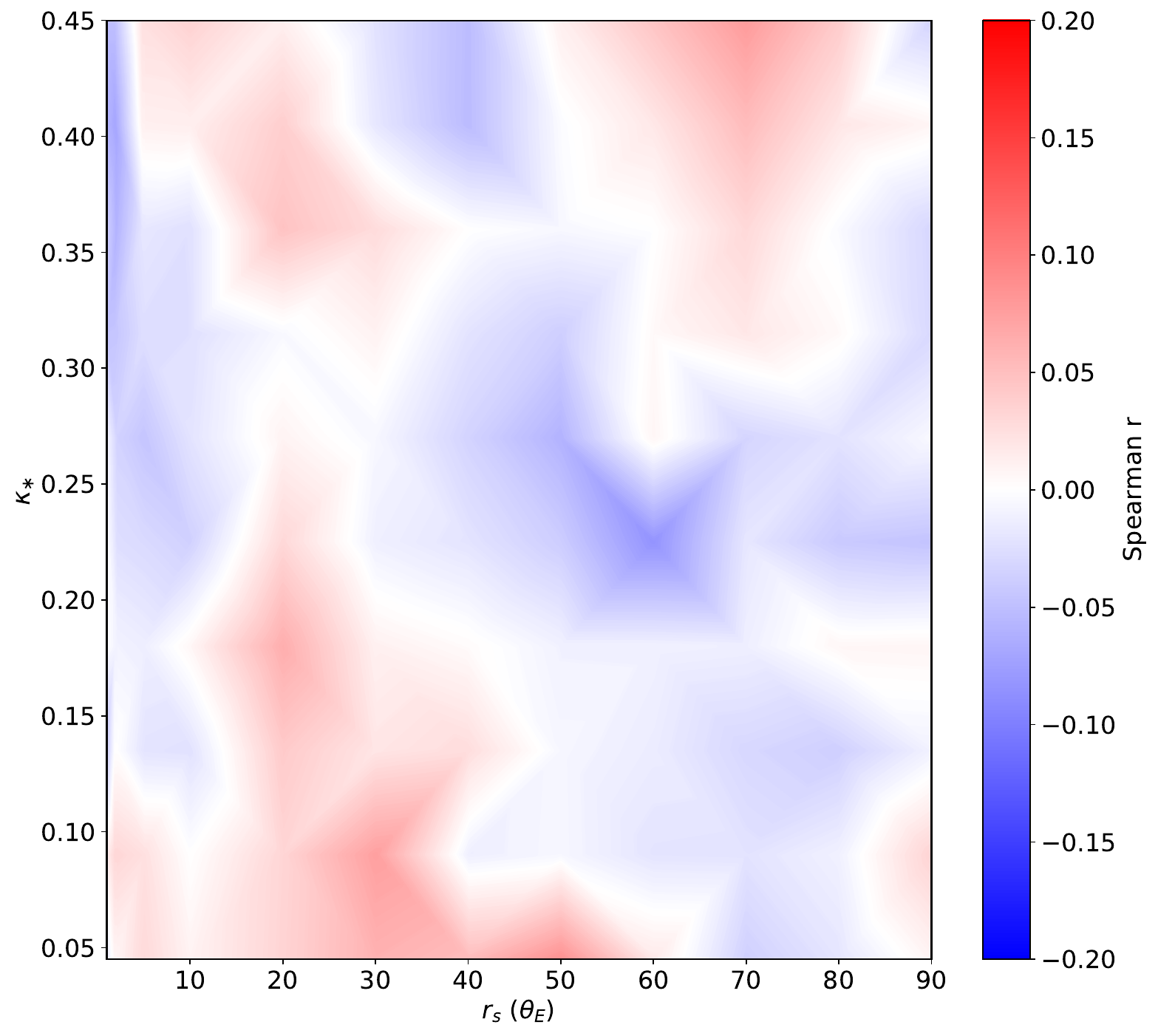}
\caption{On the left panel, the centroid shift dependence is shown for different source sizes from 1$R_E$ to 90$R_E$ with respect to the largest source of 100$R_E$ and for different convergence in microlenses $\kappa_{\ast}$ from 0.045 to 0.45. On the right panel, the Spearman correlation coefficient $r$ between magnification and shift for the $r_s$-$\kappa_{\ast}$ pairs considered is displayed. A positive (negative) correlation depicted in red (in blue) and the complete lack of correlation, $0$, is presented in white. \label{fig:shift_size}}
\end{figure*}

In the linear regime, we can extrapolate the centroid shifts in each axis in absence of macrolensing, $\Delta C_{0,1}$ and $\Delta C_{0,2}$. In order to do so, we consider two points that have the smallest $\mu_{\text{lin}}$ and, hence, are least subjected to saturation. By dividing their centroid shifts by the corresponding magnification and averaging both results, we obtain $\Delta C_{0,1}=0.681\pm0.011R_E$ and $\Delta C_{0,2}=0.698\pm0.011R_E$ which are rather similar. We can compare this centroid shift with the probability of the deflection angle for a distribution of stars, which in the regime of multiple scatterings the probability distribution function, PDF, is a Gaussian of width proportional to $\sqrt{\kappa_{\ast}}\theta_E$ \citep{1986ApJ...306....2K,2009JMP....50g2503P}. Following Equation (17) of \citet{1986ApJ...306....2K}, the PDF width is 1.16$R_E$ (for the $N=7.1\cdot10^6$ microlenses considered and rescaling it to the case of no magnification) which is the same order as the centroid shift $\Delta C_0=0.976\pm0.011R_E$. This value represents the expected value of the strength of the centroid shift of a source of 1$R_E$ for $\kappa_{\ast}=0.09$ when $\kappa=\gamma=0$ ($\mu=1$) and to obtain a quick estimation for any value of $\kappa$ and $\gamma$ within the linear regime, we only need to multiply $\Delta C_0$ by $\mu_{\text{lin}}/\sqrt{2}$ following the Equation (\ref{eq:Delta_mu}).

\subsection{Dependence of the centroid shift with the source size and the abundance of microlenses}\label{sec:gen_size-alpha}

This is the key study to support the single-epoch method to measure astrometric microlensing. We will consider several source sizes in Einstein radius units (from $r_s=1$ to $100$ $R_E$) and values of $\alpha$ (the fraction of mass in microlenses, $\alpha=\kappa_{\ast}/\kappa$) ranging from $\alpha=0.1$ to $1$. We consider a typical case of magnification $\mu=10$ with $\kappa=\gamma=0.45$. To analyze the combined effect of $\kappa_{\ast}$ and $r_s$ on the centroid shift, we present on the left panel of Figure \ref{fig:shift_size} a 2D map of the centroid shift as a function of the source size and the fraction of mass in microlenses, $\Delta C(r_s,\kappa_{\ast})$. The centroid shifts are computed with respect to the largest source, in this case $r_s=100R_E$, to check whether a large source can be used as a reference for astrometric microlensing studies.

\begin{figure*}
\includegraphics[width=\linewidth]{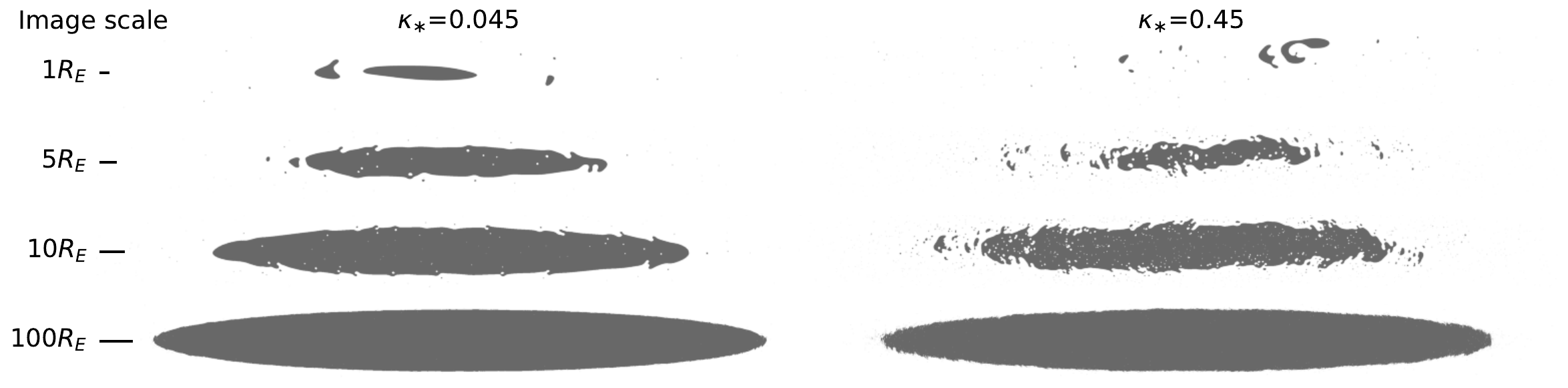}
\caption{Deformation of circular sources of radius 1, 5, 10, 50 and 100 $R_E$ due to a convergence in microlenses of 0.045 (left) and 0.45 (right). The scale of the source size radius is shown as a horizontal line on the left. \label{fig:images}}
\end{figure*}

From Figure \ref{fig:shift_size} it is clear that the centroid shifts decrease with source size and increase with stellar density. For lower values of convergence in microlenses, the slope of the decrease is flatter than for large $\kappa_{\ast}$. However, larger convergences seem to require a larger source as reference to have a centroid shift close to zero. The average image size profile of an extended source that is microlensed is a convolution of the source size with the deviations by the microlenses, i.e., $\sqrt{\sigma_W^2+\kappa_{\ast}\theta_E^2}$ where $\sigma_W$ is the source width \citep[see, e.g.,][]{1986ApJ...306....2K,2003A&A...404...83N,2021arXiv210412009D}. The profile broadening due to the source size is always symmetrical (elliptical, in fact, due to the convergence and shear) and does not contribute to the centroid shift, whereas the second term is related to the deflection of the stars by the $\kappa_{\ast}$ parameter and can produce asymmetrical profiles. From this relation, it is clear that the centroid shift will depend on both the source size and the convergence in stars. For larger $\kappa_{\ast}$ the profile can be more asymmetrical and the centroid shift will be larger and, conversely, larger sources will be imaged as large ellipses that might counteract the shift produced by the microlenses. To grasp this effect, we show in Figure \ref{fig:images} the images of several sources for the smallest and largest $\kappa_{\ast}$ considered. The stellar distribution chosen among the 1000 realizations is the one which has shifts close to the average for the smallest source. For a source of 1$R_E$, the image is quite distorted but as the size grows larger, the images become more regular as if all the mass were distributed smoothly. For the case of largest $\kappa_{\ast}$, the images are more distorted than for smaller $\kappa_{\ast}$ due to the fact that there are more stars in the vicinity of the source that affect the light rays and leads to larger shifts.

\subsection{Correlation between the centroid shifts and the flux magnification induced by microlensing}

\begin{figure*}
\includegraphics[width=\linewidth]{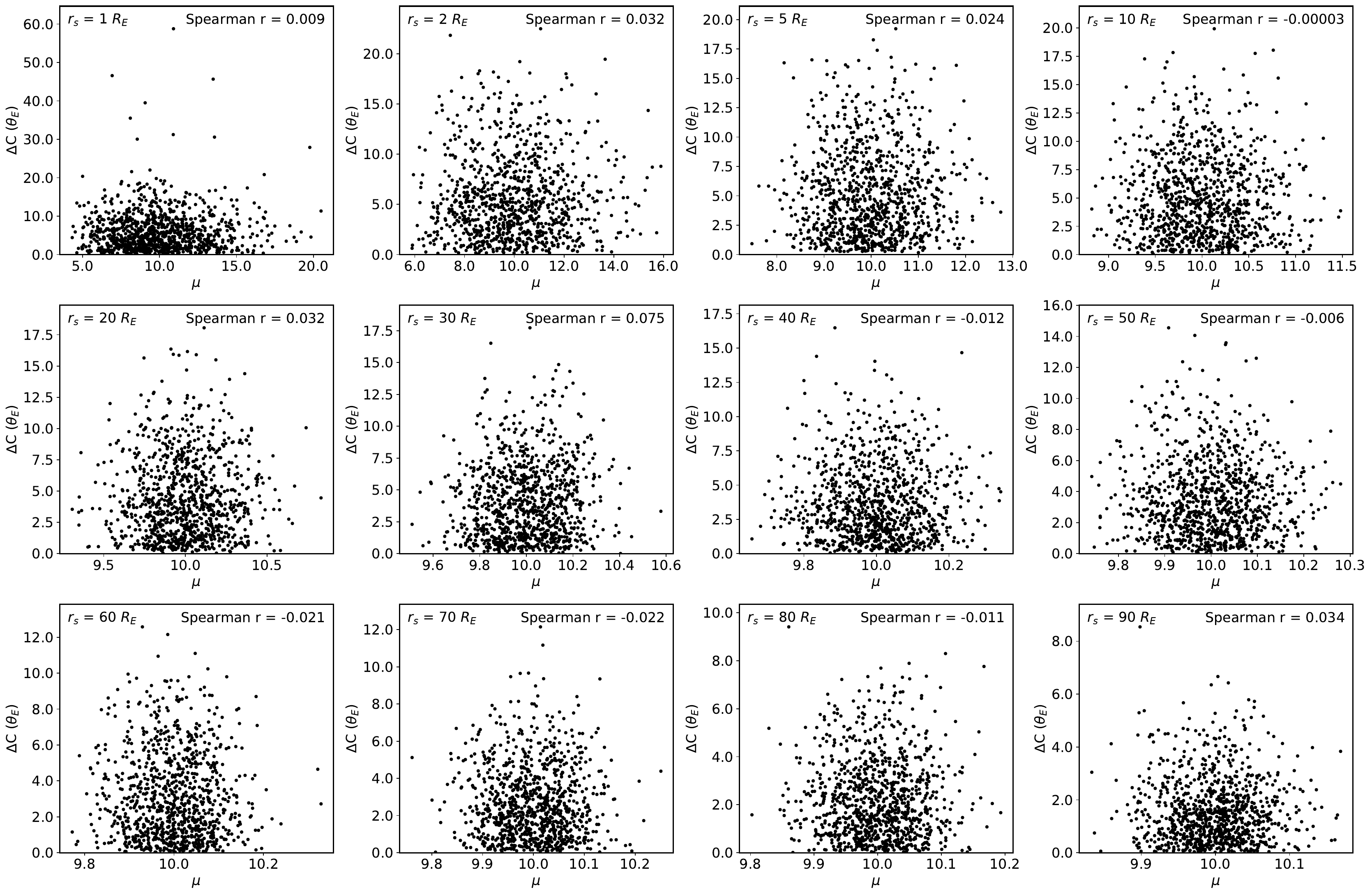}
\caption{Scatter plots of centroid shifts, $\Delta C$, and magnifications, $\mu$, for the typical case $\alpha=0.2$ ($\kappa_{\ast}=0.09$). The source size is indicated in the upper left corner of each plot and the Spearman r coefficient is in the upper right corner. \label{fig:corr}}
\end{figure*}

Since the microlensing-induced shifts are expected to be small, one possible strategy for detecting them is to observe the lens systems at epochs when large shifts are more likely to occur. To detect possible correlations for our set of simulations, we compute for every realization of the microlens distribution of the previous subsection their corresponding magnification map with the FMM-IPM code \citep{2022ApJ...941...80J} and convolve them with the source profiles. For each pair of $r_s$ and $\kappa_{\ast}$ we compute the Spearman correlation coefficient $r$ between the magnification at the source position and the centroid shift of their corresponding 1000 realizations. We choose the Spearman correlation coefficient over the Pearson correlation coefficient because the former does not assume a linear relationship between parameters, only a monotonically increasing or decreasing relationship. On the right panel of Figure \ref{fig:shift_size} we show the correlation coefficient in a 2D map of $\kappa_{\ast}$ versus $r_s$. The average correlation and its standard deviation only reaches the $-0.005\pm0.016$ level, and the maximum and minimum correlation coefficients are $+0.086$ and $-0.084$, respectively. We can argue, hence, that there is no correlation between magnification and centroid shift in random realizations of the microlens distribution. One interpretation of this finding might be that images with a large number of microimages (high magnification) are not guaranteed to be asymmetrically distributed (large shift). Also the opposite is not guaranteed, that is, low magnified images may not have their microimages symmetrically spread.

As an illustrative example of the scatter between centroid shifts and magnifications, Figure \ref{fig:corr} shows the distribution of these parameters for all sizes considered for the typical microlens mass fraction $\alpha=0.2$ ($\kappa_{\ast}=0.09$). There is no obvious correlation between the parameters but it must be noted that for sizes above 40$R_E$ the variance does not exceed more than (9.5, 10.5) in magnification. In such cases, the possible correlation between shift and magnification might be diluted and such effect must be taken into account. In our case, however, the lack of correlation obtained for smaller sizes (with the same order of magnitude as for larger sources) points out that the centroid shifts obtained from different random stellar distributions are not correlated with the magnifications irrespective of the source size.

Given the lack of correlation, the choice of high-magnification events in single-epoch measurements does not guarantee an increase of centroid shifts because of the random distributions of microlenses. \citet{1998ApJ...501..478L} found a slight correlation between these parameters 
when the source was placed at subsequent positions of the magnification map but no correlation coefficient was reported. \citet{2004A&A...416...19T} found positive correlations when the source is moving in the magnification map and specially strong correlations are found when the source traverses its own size and crosses a caustic. Hence, in differential astrometric campaigns where the position of the image is tracked during a certain period of time, large magnification events may be used as alerts to intensively measure the image position as its shift is likely to be greater. However, in single-epoch measurements, the shift with respect to an unmicrolensed reference it is not guaranteed to be larger than in low magnification events.

\begin{figure*}
\centering
\includegraphics[width=\linewidth]{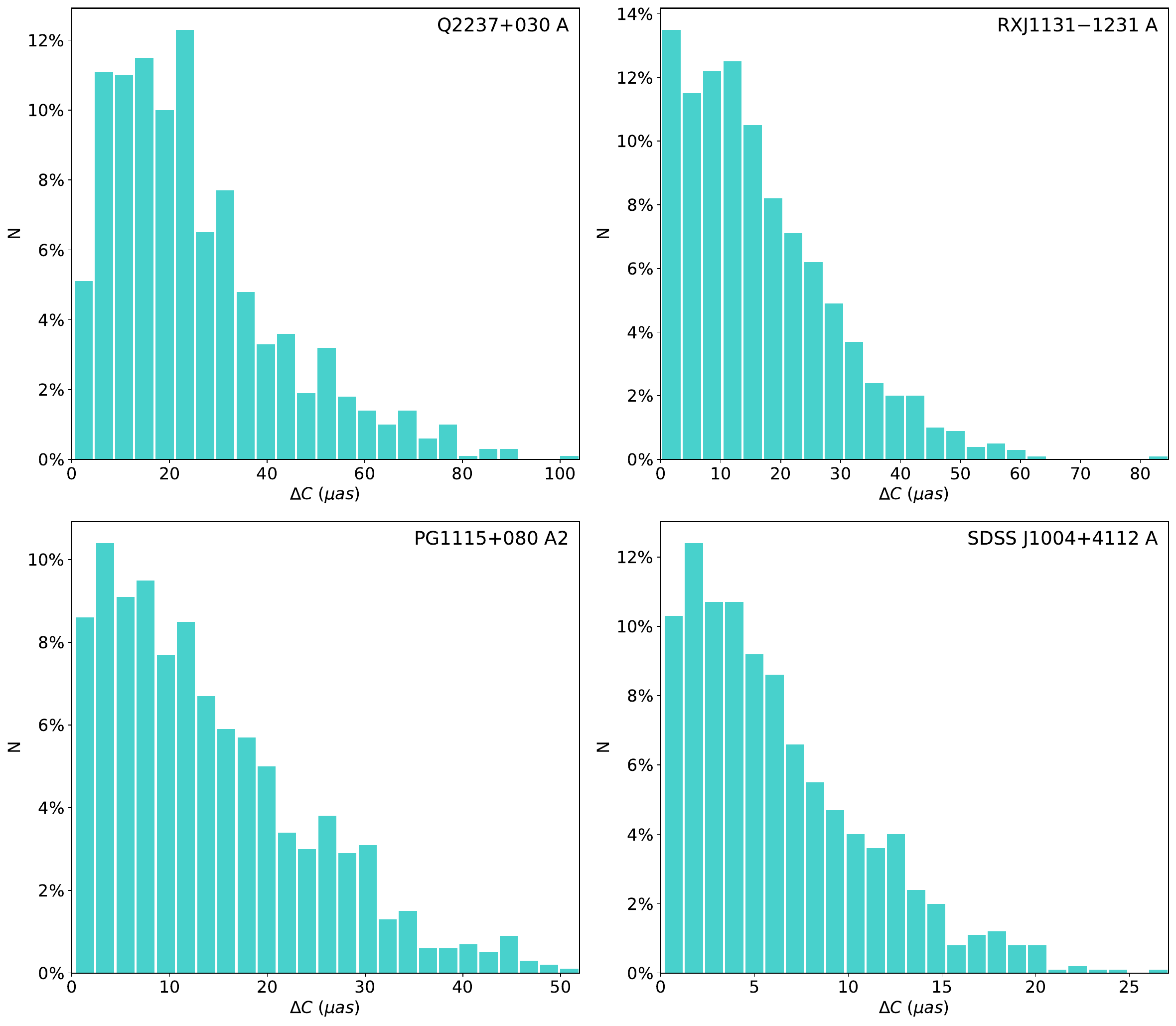}
\caption{Occurrence rate of centroid shifts for the four simulated lens systems. The upper left panel shows the shifts of a $0.3R_E$ source in the image A of the lens system Q2237+030. The upper right panel shows the shifts of a $0.06R_E$ source in the image A of the lens system RXJ1131$-$1231. The lower left panel shows the shifts of a $0.7R_E$ source in the image A2 of the lens system PG1115+080. The lower right panel shows the shifts of a $0.6R_E$ source in the image A of the lens system SDSS~J1004+4112. \label{fig:sys_hist}}
\end{figure*}

\section{Specific study of the best-suited cases}\label{sec:specific}

According to the previous simulations, the best-suited systems should be those with large abundance of microlenses, $\kappa_{\ast}$, and linear magnification, $\mu_{lin}$, but small source size, $r_s$. Also, all these results were reported in units of the Einstein angle, $\theta_E$, so the physical angular shift will be directly proportional to $\theta_E$, which in turn depends on the redshifts of the source and lens and the mean mass of microlenses. We can estimate the order of centroid shifts of specific lens systems by converting the centroid shifts of Figure \ref{fig:shift_size} in each direction in shifts in absence of macrolensing, $\Delta \vec C$. Then, we interpolate from the 2D maps the centroid shift in absence of macrolensing at the specific values of ($r_s$,$\kappa_{\ast}$) for each lens system. This can be converted into the estimation of the centroid shift by following Equation \ref{eq:Delta_vec} and multiplying it by the Einstein radius in $\mu$as. Since we need to know $\kappa$, $\kappa_{\ast}$, $\gamma$, $r_s$ and $\theta_E$, we only considered the lens systems in the GERLUMPH\footnote{http://gerlumph.swin.edu.au/} macromodel database \citep{2014ApJS..211...16V}. $r_s$ and $\theta_E$ are taken from \citet{2011ApJ...738...96M}, assuming stars with a mean mass $\langle M \rangle=0.3M_{\odot}$ as microlenses. $\kappa_{\ast}$ corresponds to $\alpha=\kappa_{\ast}/\kappa=0.2$, which is the typical stellar fraction for galaxies \citep{2015ApJ...799..149J}, unless otherwise stated. The centroid shift estimations of each image from the lens systems considered are reported in Appendix \ref{sec:appendix}. From these estimates, we select the following interesting cases to accurately simulate their centroid shifts.

\subsection{Image A of Q2237+030}\label{sec:sp_2237}

We model the lens system Q2237+030 because it has the largest centroid shift prediction in the sample. This gravitational lens was already pointed out by \citet{1998ApJ...501..478L} and \citet{2004A&A...416...19T} as an ideal case for astrometric microlensing. We simulate 1000 centroid shift realizations using $\kappa=0.4$ and $\gamma=0.4$ from the macromodel of \citet{2010ApJ...712..658P}. We consider the usual value of $\alpha=1$ for this system given that the images are located in the inner part of the lens galaxy. The source is a uniform circle of $0.3R_E$ whose half-light radius corresponds to the value obtained by \citet{2016ApJ...817..155M} shifted to the B filter central wavelength ($\lambda_{\text{obs}}=4450$\AA) using their inferred wavelength dependence of $p=0.66$. The histogram of the expected shifts for this system is shown in the upper left panel of Figure \ref{fig:sys_hist}. The centroid shifts can reach up to 100 $\mu$as with a median value of 21 $\mu$as.

\subsection{Image A of RXJ1131$-$1231}\label{sec:sp_1131}

Image A of the lens system RXJ1131$-$1231 is the fourth image with the largest predicted centroid shift but the second lens system with largest $\Delta C$. The macromodel parameters for image A are $\kappa=0.569$, $\gamma=0.465$ and $\alpha=0.11$, extracted from the mass model of \citet{2010ApJ...709..278D} with $f_{\ast}=0.3$, because this is the mass model favored by the microlensing study performed in the same work. We place a circular source profile of $0.06R_E$, which corresponds to the size at $\lambda_{\text{obs}}=4450$\AA\ \citep{2010ApJ...709..278D}, following a thin-disk wavelength dependence, $p=4/3$. In the upper right panel of Figure \ref{fig:sys_hist} half of the cases have shifts above 14 $\mu$as and the maximum shift achieved is around 60 $\mu$as.

\subsection{Image A2 of PG1115+080}\label{sec:sp_1115}

The image A2 of the lens system PG1115+080 is the next image to have large centroid shifts according to our estimations. We model the centroid shifts by using the mass model of \citet{2008ApJ...689..755M} ($\kappa$=0.61, $\gamma$=0.44 and $\alpha=0.12$) and their inferred optical size from microlensing, which favors the $f_{\ast}=0.4$ case. The circular face-on size in the center of the B-band following the thin-disk size-wavelength dependence is $0.7R_E$. The histogram of 1000 centroid shifts from different random stellar distributions is depicted in the lower left panel of Figure \ref{fig:sys_hist}. For this system, the centroid shifts can be as large as 50 $\mu$as but the median is 12 $\mu$as.

\subsection{Image A of SDSS~J1004+4112}\label{sec:sp_1004}

Despite that the image A of SDSS~J1004+4112 is the 16th image with the largest $\Delta C$, it is worth studying its centroid shifts given that photometric microlensing variability has been detected even though a galaxy cluster is acting as a lens and has a magnification much larger than most typical galaxy lenses. The centroid shifts are computed for $\kappa=0.73$ and $\gamma=0.33$ from \citet{2022ApJ...937...35F}. We adopt a stellar fraction of $\alpha=0.075$ and a circular source size of $0.6R_E$ based on the inferences in \citet[][submitted]{Fores-Toribio_subm} assuming a thin-disk wavelength dependence to estimate the source size at $\lambda_{\text{obs}}=4450$ \AA. In this case, the centroid shifts reach up to 25 $\mu$as and with a median value of 5 $\mu$as, smaller than the rest, but compatible with what was expected in Table \ref{tab:sys_shifts}.

\section{Discussion}\label{sec:disc}

It is convenient to discuss the possible applications of astrometric microlensing attending to the amplitude of the centroid shift, which depends on the mass of the microlens. For typical configurations of lens galaxies and quasar sources $|\Delta \vec C| \sim \sqrt{\mu_+^2+\mu_-^2}\sqrt{M/0.3M_\odot}\, \rm \mu as$. For linear magnifications about 10 and stellar masses of $0.3M_{\odot}$, we obtain centroid shifts of the order tens of $\mu$as (microlensing scale), while for masses above $10^3M_\odot$, we obtain shifts of the order of mas (millilensing scale). Between these scales, the mass range from $10$ to $10^2$ $M_\odot$ has special interest because it corresponds to the masses of the black holes (BHs) detected by the LIGO–Virgo–KAGRA Collaboration \citep{2023PhRvX..13d1039A}.

It should be noted that the results of the previous sections are computed for distributions of microlenses of the same mass. However, the actual mass distributions of stars or other compact objects have a mass spectrum that may alter the results to some extent. It is commonly accepted that the average mass of the distribution is the representative quantity and the results can be scaled to this mass. \citet{1986ApJ...306....2K} (Equation 24b) found that the deflection by a spectrum mass of microlenses mainly depends on $\langle m^2\rangle/\langle m\rangle$ along with the average $\kappa_{\ast}$. Hence, when a large range in the mass of the microlenses is considered, the deflections not only depend on the average mass \citep[see, e.g.,][]{2020ApJ...904..176E,2021arXiv210412009D}. For the sake of simplicity, in the following sections we will only discuss the effect of microlenses of different mass ranges separately. A joint analysis with all possible microlens masses is beyond the scope of the paper but it should be performed to properly assess the impact of each population on the astrometry.

\subsection{Stellar mass lenses, $M\sim M_\odot$ (microlensing): multiply imaged quasars}

Since there is a dependence between the observed wavelength and the size of the emitting region, the continua at different wavelengths should experiment centroid shifts of different amplitude according to our simulations. This will help to study the structure of the accretion disk, in particular its temperature profile. The same reasoning can be applied to the BLR, which is commonly accepted to be stratified according to the line ionization degree.

\subsection{Intermediate-mass lenses, $10M_\odot \lesssim M\lesssim 10^2 M_\odot$: LIGO-Virgo-KAGRA BHs}

This range of masses is very interesting because there is evidence from LIGO-Virgo-KAGRA experiments of the existence of BHs with masses in this interval. Shifts of order $10^2$ $\mu$as are expected, however, the Einstein radius of the BHs would be around $10^2$ light-days and, according to Figure \ref{fig:shift_size}, a reference source insensitive to microlensing should have a size of $10^4$ light-days, significantly exceeding the size of BLRs. An option to apply the single-epoch method above this mass might be to adopt as a reference a larger region (e.g., the infrared torus or the extended radio emission region). Alternatively, we can measure the differences between the positions of several images of a gravitationally lensed quasar at different epochs, namely a differential astrometric microlensing approach. The relative shifts between images should have a random origin and be, consequently, uncorrelated.

\subsection{Large-mass lenses, $10^3M_\odot \lesssim M\lesssim 10^6 M_\odot$ (millilensing): PMBHs, PBH clusters}

Shifts in the 0\farcs001 to 0\farcs01 range are expected in this interval of masses, which could be routinely measured with normal CCD photometry in a medium sized telescope. Comparison of astrometry from the literature of HST, JWST and/or ground based telescopes of quads taken at different epochs should be useful to put constraints on any population of compact objects in this mass interval. If a significant population of millilenses of these masses exists, it should have a strong impact on lens-model fitting, as the induced centroid shifts exceeds the uncertainties in astrometric models.

\subsection{Experimental approach}

There are several instrumental possibilities, present and future, to perform astrometric measurements with precisions similar to those required by our calculations. With the ESA's Gaia space mission, high-precision differential astrometry in the separation of relatively bright lensed quasar images with accuracies of a few $\mu$as may be obtained in the optical band \citep{2016A&A...595A...1G}. This precision can be improved with GRAVITY, which combines the near-infrared light of the four 8.2m telescope of ESO's VLT-I \citep{2021A&A...647A..59G}, or with Theia, which promise submicroarcsecond astrometric precisions. The next generation of instruments like HARMONI, designed to obtain integral field spectroscopy at the maximum spatial resolution of the E-ELT, may be used to apply the single-epoch method to a large number of quasar images.

\section{Conclusions}\label{sec:concl}

The main conclusions that can be derived from our simulations are the following:

1. The amplitude of the centroid shifts, $\Delta C$, increases with the linear magnification, $\mu_{\text{lin}}=\sqrt{\mu_-^2+\mu_+^2}$, almost linearly. Hence, the amplitude of the centroid shifts can be estimated as the product of the centroid shift in absence of macromagnification,  $|\Delta \vec C_0|$, and the linear macromagnification (see Equation \ref{eq:Delta_mu}). Theoretical analyses of the random deflections of microlenses \citep{1986ApJ...306....2K,2009JMP....50g2503P} show that the amplitude depends on the square root of the convergence in microlenses and is directly proportional to the Einstein angle $\theta_E$. The $|\Delta \vec C_0|$ found in our simulations is similar to the expected PDF width of the deflection from \citet{1986ApJ...306....2K}.

2. The average amplitude of the centroid shift increases with the convergence of microlenses and decreases with size with a gentle slope, approximating zero only for size values of around 100 Einstein radii. The dependence of the average image profile on these two parameters has been already proven \citep{2021arXiv210412009D} but in Figure \ref{fig:shift_size} the impact on the centroid shift has been computed numerically. The gentle size dependence limits the use of the BLR as reference for zero centroid shift for microlenses masses $M\lesssim 10M_\odot$. Above this mass, the BLR becomes sensitive to astrometric\footnote{We notice that in the case of flux magnification induced by microlensing the BLR becomes critically sensitive to microlensing when the Einstein radius of the microlenses is comparable to the size of the region, and this typically occurs for $M\gtrsim 100M_\odot$.} microlensing and we need larger emission regions (e.g., the infrared torus or the extended radio emission region) as reference to apply the single-epoch method or, alternatively, to consider differential measurements of the distances between images in a gravitationally lensed quasar in several epochs.

3. We have found no correlation between centroid shift and magnification (macromagnification plus micromagnification) between random distributions of microlenses. Hence, a single-epoch astrometric observation during a high-magnification event does not guarantee finding large centroid shift with respect to the baseline. However, in the differential astrometry approach, a correlation arises \citep{1998ApJ...501..478L,2004A&A...416...19T} due to the tracking during subsequent source positions.

4. We select four gravitational lensed images of quasars which will be prone to astrometric microlensing and study the centroid displacements. The image A of Q2237+030 is the most favorable system with a 50\% chance of measuring centroid shifts larger than 21 $\mu$as for a source size of 0.3$R_E$ and microlens masses of 0.3$M_{\odot}$.

5. Strategy for measurements: We propose to use spectro-astrometry in upcoming integral field units to simultaneously obtain single-epoch photocenters of the continuum and of different emission line regions when probing microlenses of stellar masses and stratification of quasar accretion disks. If a large emission region for reference is not available, then a differential astrometry approach can be applied between multiple images of lensed quasars in different epochs with high-precision instrumentation.

\begin{acknowledgments}
We thank the referee for their comments and suggestions, which have significantly improved this work. This research was supported by the grants PID2020-118687GB-C31, PID2020-118687GB-C32 and PID2020-118687GB-C33 financed by the Spanish Ministerio de Ciencia e Innovaci\'on through MCIN/AEI/10.13039/501100011033. J.A.M. is also supported by the Generalitat Valenciana with the project of excellence Prometeo/2020/085. J.J.V. is also supported by the project FQM-108, financed by Junta de Andaluc\'{\i}a. CdB acknowledges support from a Beatriz Galindo senior fellowship (BG22/00166) from the Spanish Ministry of Science, Innovation and Universities.
\end{acknowledgments}

\begin{appendix}
\section{Estimated centroid shifts for specific lens systems}\label{sec:appendix}

The expected centroid shifts following the procedure described in \S\ref{sec:specific} are presented in Table \ref{tab:sys_shifts} in decreasing order of $\Delta C$. The sources radii in \citet{2011ApJ...738...96M} are estimated using their I-band magnitudes and a thin-disk model \citep{1973A&A....24..337S}. This method typically underestimates the source sizes in comparison to estimations from black hole masses and luminosities using the same thin-disk theory as stated in \citet{2011ApJ...738...96M}. To obtain a safe estimate of the centroid shifts, we rescale the quasar sizes to a median value of $5$ light-days as the typical size obtained with the latter scaling relation (see e.g. \citet{2020ApJS..246...16Y} for a black hole mass of $10^9M_{\sun}$ and an Eddington ratio of $\dot{m}_E=0.3$). 
\startlongtable
\begin{deluxetable*}{l l D D D D D D c}

\tablecaption{Estimation of centroid shifts, $\Delta C$, for 79 lensed quasar images in the GERLUMPH database \citep{2014ApJS..211...16V}.}
\label{tab:sys_shifts}
\tablehead{
Lens system & Image & \multicolumn2c{$\kappa$} & \multicolumn2c{$\gamma$} & \multicolumn2c{$\alpha$} & \multicolumn2c{$R_{1/2}$ ($\theta_E$)} & \multicolumn2c{$\theta_E$ ($\mu$as)} & \multicolumn2c{$\Delta C$ ($\mu$as)} & References 
}
\decimals
\startdata
Q2237+030 & A      & 0.4 & 0.4 & 1.0 & 1.05 & 3.90 & 21.17 & (5) \\ 
Q2237+030 & D      & 0.62 & 0.62 & 1.0 & 1.05 & 3.90 & $\gtrsim$19.24 (14.15)\tablenotemark{a} & (5) \\ 
Q2237+030 & B      & 0.38 & 0.39 & 1.0 & 1.05 & 3.90 & 17.93 & (5) \\ 
RXJ1131$-$1231 & A   & 0.569 & 0.465 & 0.11 & 0.94 & 1.19 & 14.01 & (4) \\ 
PG1115+080 & A2    & 0.61 & 0.44 & 0.12 & 1.20 & 1.43 & 12.00 & (2) \\ 
PG1115+080 & A1    & 0.59 & 0.36 & 0.11 & 1.20 & 1.43 & 11.17 & (2) \\ 
Q2237+030 & C      & 0.73 & 0.72 & 1.0 & 1.05 & 3.90 & $\gtrsim$10.84   (6.22)\tablenotemark{a} & (5) \\ 
HE0047$-$1756 & A    & 0.45 & 0.48 & 0.2 & 1.47 & 1.23 & 8.15 & (8) \\ 
MG0414+0534 & A2   & 0.53 & 0.524 & 0.2 & 0.34 & 0.85 & 8.01 & (7) \\ 
SDSSJ0924+0219 & A & 0.39 & 0.54 & 0.2 & 0.71 & 1.24 & 7.90 & (9) \\ 
MG0414+0534 & A1   & 0.489 & 0.454 & 0.2 & 0.34 & 0.85 & 7.36 & (7) \\ 
RXJ1131$-$1231 & B   & 0.53 & 0.41 & 0.099 & 0.94 & 1.19 & 7.29 & (4) \\ 
SDSSJ0924+0219 & D & 0.4 & 0.68 & 0.2 & 0.71 & 1.24 & 6.99 & (9) \\ 
RXJ1131$-$1231 & C   & 0.546 & 0.387 & 0.103 & 0.94 & 1.19 & 6.77 & (4) \\ 
HE0230$-$2130 & B    & 0.51 & 0.587 & 0.2 & 0.72 & 1.12 & 5.77 & (7) \\
SDSSJ1004+4112 & A & 0.73 & 0.33 & 0.075 & 1.07 & 0.93 & 5.65 & (10) \\
B1422+231 & B      & 0.492 & 0.628 & 0.2 & 2.69 & 1.44 & 5.59 & (7) \\ 
HE0230$-$2130 & A    & 0.472 & 0.416 & 0.2 & 0.72 & 1.12 & 4.87 & (7) \\  
SDSSJ1138+0314 & D & 0.523 & 0.614 & 0.2 & 0.54 & 1.23 & 4.57 & (7) \\ 
SDSSJ0924+0219 & C & 0.44 & 0.7 & 0.24 & 0.71 & 1.24 & 4.53 & (9) \\ 
HE0435$-$1223 & B    & 0.539 & 0.602 & 0.2 & 0.95 & 1.16 & 4.20 & (7) \\ 
SDSSJ1138+0314 & A & 0.465 & 0.384 & 0.2 & 0.54 & 1.23 & 3.99 & (7) \\ 
RXJ0911+0551 & B   & 0.586 & 0.281 & 0.2 & 0.94 & 0.97 & 3.91 & (7) \\ 
B1422+231 & A      & 0.38 & 0.473 & 0.2 & 2.69 & 1.44 & 3.90 & (7) \\ 
SDSSJ1206+4332 & A & 0.43 & 0.41 & 0.2 & 0.84 & 1.23 & 3.60 & (3) \\ 
HE0435$-$1223 & C    & 0.444 & 0.396 & 0.2 & 0.95 & 1.16 & 3.44 & (7) \\ 
QJ0158$-$4325 & A    & 0.54 & 0.28 & 0.15 & 1.74 & 1.36 & 3.32 & (1) \\ 
HE0435$-$1223 & A    & 0.445 & 0.383 & 0.2 & 0.95 & 1.16 & 3.21 & (7) \\ 
SDSSJ0924+0219 & B & 0.38 & 0.45 & 0.19 & 0.71 & 1.24 & 3.17 & (9) \\ 
HE1104$-$1805 & A    & 0.64 & 0.52 & 0.2 & 3.77 & 0.96 & 3.15 & (3) \\ 
QJ0158$-$4325 & B    & 0.89 & 0.57 & 0.32 & 1.74 & 1.36 & 3.10 & (1) \\ 
RXJ0911+0551 & A   & 0.646 & 0.544 & 0.2 & 0.94 & 0.97 & 2.87 & (7) \\ 
SDSSJ1138+0314 & C & 0.438 & 0.349 & 0.2 & 0.54 & 1.23 & 2.79 & (7) \\ 
PG1115+080 & B     & 0.69 & 0.58 & 0.15 & 1.20 & 1.43 & 2.73 & (2) \\ 
SDSSJ1138+0314 & B & 0.578 & 0.673 & 0.2 & 0.54 & 1.23 & 2.66 & (7) \\ 
HE0435$-$1223 & D    & 0.587 & 0.648 & 0.2 & 0.95 & 1.16 & 2.66 & (7) \\ 
RXJ0911+0551 & C   & 0.637 & 0.577 & 0.2 & 0.94 & 0.97 & 2.54 & (7) \\ 
H1413+117 & A      & 0.53 & 0.64 & 0.2 & 1.86 & 0.86 & 2.51 & (3) \\ 
HE0512$-$3329 & A    & 0.59 & 0.55 & 0.2 & 3.76 & 0.69 & 2.47 & (3) \\ 
HE0047$-$1756 & B    & 0.62 & 0.66 & 0.2 & 1.47 & 1.23 & 2.41 & (8) \\ 
HE0230$-$2130 & C    & 0.44 & 0.334 & 0.2 & 0.72 & 1.12 & 2.40 & (7) \\ 
SDSSJ1155+6346 & B & 1.67 & 1.47 & 0.2 & 1.53 & 1.92 & 2.28 & (8) \\ 
B1422+231 & C      & 0.365 & 0.378 & 0.2 & 2.69 & 1.44 & 2.22 & (7) \\ 
PG1115+080 & C     & 0.54 & 0.25 & 0.08 & 1.20 & 1.43 & 2.19 & (2) \\ 
RXJ1131$-$1231 & D   & 1.001 & 0.635 & 0.242 & 0.94 & 1.19 & 2.19 & (4) \\ 
H1413+117 & C      & 0.46 & 0.35 & 0.2 & 1.86 & 0.86 & 2.13 & (3) \\ 
WFI2033$-$4723 & A1  & 0.506 & 0.255 & 0.2 & 1.10 & 0.93 & 2.02 & (7) \\ 
SDSSJ1206+4332 & B & 0.63 & 0.72 & 0.2 & 0.84 & 1.23 & 2.02 & (3) \\ 
MG0414+0534 & B    & 0.46 & 0.316 & 0.2 & 0.34 & 0.85 & 1.89 & (7) \\ 
SDSSJ1004+4112 & B & 0.65 & 0.23 & 0.048 & 1.07 & 0.93 & 1.86 & (10) \\ 
Q1017$-$207 & A      & 0.35 & 0.45 & 0.2 & 3.21 & 0.95 & 1.83 & (3) \\ 
WFI2033$-$4723 & A2  & 0.665 & 0.643 & 0.2 & 1.10 & 0.93 & 1.76 & (7) \\ 
SDSSJ1001+5027 & B & 0.74 & 0.72 & 0.2 & 1.91 & 1.22 & 1.71 & (3) \\ 
H1413+117 & B      & 0.43 & 0.34 & 0.2 & 1.86 & 0.86 & 1.71 & (3) \\ 
H1413+117 & D      & 0.58 & 0.69 & 0.2 & 1.86 & 0.86 & 1.67 & (3) \\ 
SDSSJ1353+1138 & B & 0.96 & 0.89 & 0.2 & 2.80 & 1.44 & 1.53 & (3) \\ 
FBQ0951+2635 & B   & 1.07 & 1.02 & 0.2 & 3.67 & 1.57 & 1.49 & (3) \\ 
HE0230$-$2130 & D    & 1.07 & 0.864 & 0.2 & 0.72 & 1.12 & 1.44 & (7) \\ 
WFI2033$-$4723 & B   & 0.392 & 0.302 & 0.2 & 1.10 & 0.93 & 1.41 & (7) \\ 
MG0414+0534 & C    & 0.676 & 0.693 & 0.2 & 0.34 & 0.85 & 1.40 & (7) \\ 
SDSSJ1001+5027 & A & 0.35 & 0.28 & 0.2 & 1.91 & 1.22 & 1.37 & (3) \\ 
WFI2033$-$4723 & C   & 0.7 & 0.735 & 0.2 & 1.10 & 0.93 & 1.34 & (7) \\ 
Q0957+561 & B      & 1.03 & 0.91 & 0.2 & 3.51 & 1.28 & 1.34 & (3) \\ 
SDSSJ0806+2006 & B & 0.82 & 0.77 & 0.2 & 1.16 & 1.04 & 1.33 & (3) \\ 
HE0512$-$3329 & B    & 0.41 & 0.37 & 0.2 & 3.76 & 0.69 & 1.32 & (3) \\ 
SDSSJ0806+2006 & A & 0.34 & 0.29 & 0.2 & 1.16 & 1.04 & 1.22 & (3) \\ 
SDSSJ1353+1138 & A & 0.3 & 0.22 & 0.2 & 2.80 & 1.44 & 1.19 & (3) \\ 
FBQ0951+2635 & A   & 0.28 & 0.15 & 0.2 & 3.67 & 1.57 & 1.12 & (3) \\
SDSSJ1004+4112 & C & 0.59 & 0.23 & $<$0.045 & 1.07 & 0.93 & $<$1.09 & (10) \\ 
SDSSJ1155+6346 & A & 0.22 & 0.03 & 0.2 & 1.53 & 1.92 & 1.09 & (8) \\ 
SDSSJ1004+4112 & D & 1.02 & 0.49 & 0.066 & 1.07 & 0.93 & 1.08 & (10) \\
SBS1520+530 & B    & 0.9 & 0.85 & 0.2 & 2.80 & 0.91 & 1.00 & (3) \\ 
SBS0909+523 & A    & 0.66 & 0.77 & 0.2 & 9.12 & 0.71 & 0.88 & (6) \\ 
HE1104$-$1805 & B    & 0.33 & 0.21 & 0.2 & 3.77 & 0.96 & 0.86 & (3) \\ 
Q1017$-$207 & B      & 1.23 & 1.32 & 0.2 & 3.21 & 0.95 & 0.79 & (3) \\ 
RXJ0911+0551 & D   & 0.29 & 0.066 & 0.2 & 0.94 & 0.97 & 0.74 & (7) \\ 
SBS1520+530 & A    & 0.29 & 0.15 & 0.2 & 2.80 & 0.91 & 0.68 & (3) \\ 
Q0957+561 & A      & 0.2 & 0.15 & 0.2 & 3.51 & 1.28 & 0.67 & (3) \\ 
SBS0909+523 & B    & 0.36 & 0.25 & 0.2 & 9.12 & 0.71 & 0.67 & (6) \\  
\enddata
\tablecomments{The macromodels used ($\kappa$, $\gamma$ and $\alpha$) are shown along with their references. The half-light radius, $R_{1/2}$, and the Einstein angle, $\theta_E$, are taken from \citet{2011ApJ...738...96M}.}
\tablenotetext{a}{Lower level estimations given that their $\kappa_{\ast}$ are larger than the maximum $\kappa_{\ast}$ in Figure \ref{fig:shift_size}. The result using the PDF width of random microlensing deflections of \citet{1986ApJ...306....2K} with $N=10^6$ microlenses as an estimate of $\Delta C_0$ is reported in parentheses.}
\tablerefs{(1) $\kappa$, $\gamma$ and $\alpha$ from \citet{2008ApJ...676...80M}, $f_{M/L}=0.4$ mass model ($\alpha$ closest to 0.2). (2) $\kappa$, $\gamma$ and $\alpha$ from \citet{2008ApJ...689..755M}, $f_{M/L}=0.4$ mass model favored by microlensing. (3) \citet{2009ApJ...706.1451M}, (4) $\kappa$, $\gamma$ and $\alpha$ from \citet{2010ApJ...709..278D}, $f_{M/L}=0.3$ mass model favored by microlensing. (5) $\kappa$ and $\gamma$ from \citet{2010ApJ...712..658P} and $\alpha$ from, e.g., \citet{2016ApJ...817..155M}. (6) \citet{2011ApJ...730...16M}. (7) \citet{2014ApJ...793...96S}. (8) \citet{2014ApJ...797...61R}. (9) $\kappa$, $\gamma$ and $\alpha$ from \citet{2015ApJ...806..258M}, $f_{M/L}=0.6$ mass model ($\alpha$ closest to 0.2). (10) $\kappa$ and $\gamma$ from \citet{2022ApJ...937...35F} and $\alpha$ from \citet[][submitted]{Fores-Toribio_subm}.}
\end{deluxetable*}

\end{appendix}

\bibliographystyle{aasjournal}
\bibliography{refs}{}

\end{document}